# Two-Dimensional Multiferroics: Ferroelasticity, Ferroelectricity, Domain Wall, and Potential Mechano-Opto-Electronic Applications

Hua Wang[1], Xiaofeng Qian[1]*

[1] *Department of Materials Science and Engineering, Dwight Look College of Engineering and College of Science, Texas A&M University, College Station, Texas 77843, USA*

*Correspondence and requests for materials should be addressed to X.Q. (email: feng@tamu.edu).

**Low-dimensional multiferroic materials hold great promises in miniaturized device applications such as nanoscale transducers, actuators, sensors, photovoltaics, and nonvolatile memories. Here, using first-principles theory we predict that two-dimensional (2D) monolayer Group IV monochalcogenides including GeS, GeSe, SnS, and SnSe are a class of 2D semiconducting multiferroics with strongly coupled giant in-plane spontaneous ferroelectric polarization and spontaneous ferroelastic lattice strain that are thermodynamically stable at room temperature and beyond, and can be effectively modulated by elastic strain engineering. Their optical absorption spectra exhibit strong in-plane anisotropy with visible-spectrum excitonic gaps and sizable exciton binding energies, rendering the unique characteristics of low-dimensional semiconductors. More importantly, the predicted low domain wall energy and small migration barrier together with the coupled multiferroic order and anisotropic electronic structures suggest their great potentials for tunable multiferroic functional devices by manipulating external electrical, mechanical, and optical field to control the internal responses, and enable the development of four device concepts including 2D ferroelectric memory, 2D ferroelastic memory, and 2D ferroelastoelectric nonvolatile photonic memory as well as 2D ferroelectric excitonic photovoltaics.**

**Introduction**

Two-dimensional (2D) materials (*1-6*) research has led to the remarkable discoveries of 2D metals, semiconductors, and insulators and, equally important, a variety of rich physics such as Dirac fermions (*3*) and topological insulators (*7, 8*). Recently, coupled physical properties of 2D materials attract a lot of attention for their potentials device and energy applications. For example, monolayer BN (*9*), MoS$_2$ (*10-12*) and Group IV monochalcogenides (*13, 14*) were recently found to possess large piezoelectricity (*15*) which would allow efficient mechanical-to-electrical energy conversion. Monolayer CrSiTe$_3$ was theoretically predicted (*16*) and experimentally synthesized and demonstrated (*17*) with 2D ferromagnetic ordering, paving an avenue towards 2D spintronics. Interestingly, Peierls-distorted 1T' transition metal chalcogenide have been predicted to exhibit both quantum spin Hall effect with a nontrivial Z$_2$ topological index (*18, 19*) and ferroelasticity (*20*), suggesting the possibility of controlling the anisotropy of the topologically-protected edge states via elastic strain engineering.

2D materials mentioned above usually possess one ferroic order only. Undoubtedly, 2D multiferroic materials that hold simultaneously two or more primary ferroic (*i.e.*, ferroelectric, ferromagnetic, and ferroelastics) orders are highly desirable as they may open up unprecedented opportunities for both scientific and technological endeavors (*21-27*). Ideal multiferroics, however, require different order parameters to be strongly coupled, which in tandem with appropriate kinetic barrier of phase transition will allow facile switching of electric polarization, magnetization, and lattice strain through external disparate fields. Though highly valuable, perfect multiferroic materials especially in low dimensions are scarce, largely due to the stringent symmetry and chemistry requirements for the coexistence of two or more coupled ferroic orders and good thermodynamic stability for practical applications at room temperature (*21, 25*).

Based on first-principles theory, here we predict that monolayer Group IV monochalcogenides (including GeS, GeSe, SnS, and SnSe) represent a class of 2D multiferroic materials that simultaneously possess strongly-coupled ferroelectric and ferroelastic orders and, more importantly, have low domain wall energy and small migration barrier. The calculated giant intrinsic in-plane electric polarization implies that their thickness is not constrained by out-of-plane depolarization-induced instability that often occurs in free-standing multiferroic ultrathin films(*28*). As ferroelastic order in these monolayers is pertinent to spontaneous strain of unit cell along two perpendicular orientations, the direct coupling between ferroelastic lattice strain and ferroelectric polarization allows the direct control of one ferroic order by applying external field that is conjugated with the other one. Furthermore, the spontaneous lattice strain and in-plane polarization also lead to highly anisotropic electronic and optical properties. Therefore, these 2D ferroelastoelectric semiconducting monolayers with highly anisotropic, strongly coupled, and externally switchable physical properties will engender a wide variety of ultrathin mechano-opto-electronic applications. Applications of elastic strain, electric field, or optical field can efficiently switch the multiferroic states and alter their electric, optical, and mechanical responses, thereby enabling the conceptual designs of 2D ferroelectric, ferroelastic, and ferroelastoelectric nonvolatile photonic memories as well as 2D ferroelectric excitonic photovoltaics presented below.

**Results**

Monolayer Group IV monochalcogenide (abbreviated as MX), as shown in Fig. 1A, consists of two puckered atomic layers similar to monolayer black phosphorus where M=(Ge, Sn) and X=(S, Se). We use GeSe as an example and set the *x*-axis and *y*-axis as in-plane axes with the *z*-axis along the plane normal. Its noncentrosymmetric unit cell is illustrated by dashed orange rectangle in Figs. 1B-E, which contains four atoms and belongs to space group Pmn2$_1$ with a mirror symmetry only ($M_y$: $y \rightarrow -y$). The lattice parameters are optimized by first-principles density functional theory (DFT) (*29, 30*) calculations (listed in Table S1), in good agreement with other theoretical results (*13, 14*). The calculated interlayer cohesive energies are almost constant, about 33-34 meV/Å$^2$ for all four MX monolayers (see Table S2), which are only slightly higher than that of graphene (~30 meV/Å$^2$) with the same PBE exchange-correlation functional and optB88-vdW correlation functional. Thus it is possible to achieve their monolayers via mechanical exfoliation. Figure 1B/1C and 1D/1E show the corresponding ferroelectric state with spontaneous polarization along -x/+x and –y/+y, respectively, while a direct comparison of Figure 1B/1C and Figure 1D/1E reveals the ferroelastic order with spontaneous strain along x and y, respectively. Below we will separately discuss the ferroelastic and ferroelectric order in monolayer MX.

Ferroelastic order in MX monolayers originates from the fact that their centrosymmetric parent phase can undergo spontaneous relaxation along both x and y direction, resulting in two different orientations of spontaneous lattice strain that are perpendicular to each other. For 2D materials, it can be mathematically described by 2×2 in-pane transformation strain matrix using the paraelastic structure as a reference state. For one of the two ferroelastic ground states of monolayer GeSe with spontaneous tensile strain along x, the lattice parameters $a$ and $b$ are 4.26 and 3.98 Å, respectively, which are 4.10 and 4.10 Å in the reference paraelastic state. According to its space group $Pmn2_1$, two in-plane lattice vectors are perpendicular to each other. Hence, the corresponding 2×2 in-plane unit cells $\mathbf{H}_x$ and $\mathbf{H}_{\text{ref}}$ can be expressed by: $\mathbf{H}_x = \begin{bmatrix} 4.26 & 0 \\ 0 & 3.98 \end{bmatrix}$ and $\mathbf{H}_{\text{ref}} = \begin{bmatrix} 4.10 & 0 \\ 0 & 4.10 \end{bmatrix}$. Transformation strain matrix $\boldsymbol{\eta}_x$ can be defined using the Green-Lagrange strain tensor, $\boldsymbol{\eta}_x = \frac{1}{2}([\mathbf{H}_{\text{ref}}^{-1}]^T \mathbf{H}_a^T \mathbf{H}_a \mathbf{H}_{\text{ref}}^{-1} - \mathbf{I})$,(19) where $\mathbf{T}$ denotes matrix transpose, and $\mathbf{I}$ is a 2×2 identity matrix. For GeSe, $\boldsymbol{\eta}_x = \begin{bmatrix} 0.041 & 0 \\ 0 & -0.027 \end{bmatrix}$, corresponding to 4.1% tensile strain along x and 2.7% compressive strain along y. Similarly, we obtained transformation strain matrix $\boldsymbol{\eta}_y$ for the other ferroelastic ground state: $\boldsymbol{\eta}_y = \begin{bmatrix} -0.027 & 0 \\ 0 & 0.041 \end{bmatrix}$, indicating a 2.7% compressive strain along *x* and 4.1% tensile strain along *y*. The results for other materials can be found in Table S3.

To have a better physical picture of the ferroelastic order, we calculate the total energy of fully relaxed monolayer GeSe as function of lattice parameters *a* and *b* with respect to the ground-state state energy. The results are shown in Fig. 2A. Two degenerate ground states are marked by two purple dots characterized by spontaneous strain along *x* and *y*, respectively. Under relatively large in-plane lattice parameters, two ferroelastic phases are separated by a phase boundary, whereas at smaller lattice parameters they undergo a ferroelectric-to-paraelectric transition. Furthermore, generalized solid-state nudged elastic band (NEB) calculations (*31, 32*) were performed to investigate the coherent structural transition from one ferroelastic state to the other. The calculated minimum energy pathway (MEP) is shown in Fig. 2B and marked by white stars in Fig. 2A, and the saddle point is labeled by red dot. The small energy barrier of 19 meV/unit cell in monolayer GeSe highlights the possibility of fast switching upon external mechanical stress despite the fact that such switching is generally mediated by domain wall motion. The transition barriers for three other materials are listed in Table S4. It is worthy to note that the ferroelastic transition does not go through the centrosymmetric paraelectric state where Ge and Se atoms overlap exactly in the 2D plane. Instead, as shown in Fig. 2B, the Ge and Se atoms rotate around each other.

Ferroelastic order does not necessarily correlate with ferroelectric order. Monolayer black phosphorus is a good example which shares a similar structure with monolayer MX, hence it is a 2D ferroelastic material. Nonetheless, due to single phosphorus element no ferroelectric order is expected. In contrast, the two distinct chemical elements in monolayer MX give rise to appreciable difference in electronegativity and large relative displacement, consequently they are very likely to have large spontaneous polarization. To confirm the above speculation, we need to choose an adiabatic pathway for the ferroelectric transition, and calculate its spontaneous polarization using the Berry phase approach based on the Kohn-Sham wavefunctions from first-principles DFT calculations. First, we scanned the full potential energy surface by fixing the Ge atoms in the unit cell, shifting Se atoms with respect to their hypothetic centrosymmetric position in the *x-y* plane, and then relaxing the atoms along the *z*-axis. Here, the lattice parameters are constrained to the initial ground state, and the centrosymmetric state serves as the reference point of total energy. Figure 3A clearly shows that two ferroelectric ground states are located at the minima of two potential energy wells. Our NEB calculations show that these two energy minima are connected by two MEPs (marked with green lines) which deviate away from the central paraelectric state, implying that bulk ferroelectric transition is realized by the relative rotation of Ge and Se atoms around their paraelectric positions, rather than by a straight translation between two ferroelectric states (marked as red line). The calculated minimum energy barrier for monolayer MXs (Table S5) spans a wide range, *i.e.* about 7, 33, 95, and 464 meV/unit cell for SnSe, SnS, GeSe, and GeS, respectively, suggesting a possibility to fine-tune the kinetic barriers of ferroelectric transition via stoichiometric controls of M and X chemical elements for device applications.

The adiabatic path through MX's centrosymmetric paraelectric state allows us to calculate the total polarization using modern theory of polarization based on the Berry phase approach (*33, 34*). Mathematical expression of spontaneous polarization $P_s$ in Wannier representation is given by

$$P_\text{s} = P^\text{f} - P^\text{i} = \frac{1}{\Omega}\sum_j\left(q^\text{f}\mathbf{r}^\text{f} - q^\text{i}\mathbf{r}^\text{i}\right) - \frac{2ie}{(2\pi)^3}\sum_n^\text{occ}\left[\int_\text{BZ} d^3\mathbf{k}\, e^{-i\mathbf{k}\cdot\mathbf{R}}\left(\left\langle u_{n\mathbf{k}}^\text{f}\left|\frac{\partial u_{n\mathbf{k}}^\text{f}}{\partial \mathbf{k}}\right.\right\rangle - \left\langle u_{n\mathbf{k}}^\text{i}\left|\frac{\partial u_{n\mathbf{k}}^\text{i}}{\partial \mathbf{k}}\right.\right\rangle\right)\right],$$

where "i" and "f" refer to the initial and final positions, and $u_{n\mathbf{k}}(\mathbf{r})$ is the periodic part of Bloch wave functions $u_{n\mathbf{k}}(\mathbf{r}) = e^{i\mathbf{k}\cdot\mathbf{r}}\psi_{n\mathbf{k}}(\mathbf{r})$, $\Omega$ is the volume of the unit cell, and the integral is performed over the first Brillouin zone. Here, the initial state is the centrosymmetric structure, and the final state is the polarized one. The above formula has taken into account both ionic and electronic contributions to the total polarization. As shown in Fig. 3B and Fig. 3C, monolayer GeSe has a large spontaneous polarization of 357.0 *pC/m*, rendering an effective bulk polarization of 35.7 $\mu C/cm^2$ if we assume an approximate layer thickness of 1nm. The latter includes both van der Waals distance and the intrinsic thickness of monolayer GeSe. Here, the spontaneous polarization was rigorously determined by identifying the continuous evolution of total polarization as a function of normalized displacement through the centrosymmetric paraelectric state, as shown in Fig. 3C for monolayer GeSe. The similar plots for the other three materials can be found in fig. S1, and the lines with different colors in Fig. 3C are shifted by multiple polarization quanta $_q$. Spontaneous polarizations for GeS, SnS, and SnSe monolayers are 484, 260, and 181 *pC/m*, respectively, corresponding to effective polarization of 48.4, 26.0, and 18.1 $\mu C/cm^2$ (see Table S6), which agrees well with a very recent work by Fei *et al.*(*35*). Monolayer GeS owns the largest spontaneous polarization owing to a sizable electronegativity difference of the two elements and, more importantly, the largest displacement between Ge and S atoms with respect to the paraelectric state.

Furthermore, as shown in Fig. 3B, the total energy as function of spontaneous polarization along the adabatic path (the red line in Fig. 3A) exhibits a characteristic double-well potential of ferroelectrics. The solid curve in Fig. 3B is obtained by fitting the energy as function of polarization in a sixth-order polynomial based on the Landau-Devonshire theory of

ferroelectrics(*36*), $E(P) = \frac{1}{2}aP^2 + \frac{1}{4}bP^4 + \frac{1}{6}cP^6$, where $E$ is the total energy with respect to the paraelectric state and $P$ is the total polarization. By including the energy contribution from the conjugate external electric field $\mathcal{E}$ and internal polarization, one can define electric enthalpy $F = E - \mathcal{E}P$, from which we can estimate the ideal coercive field $\mathcal{E}_c$ from the maximum slope of $E(P)$, *i.e.* $\mathcal{E}_c = \max\left(\frac{dE}{dP}\right)$, between the minimum and saddle point (*27*). The calculated ideal coercive field $\mathcal{E}_c$ for monolayer GeSe is 0.623 V/nm under stress-free condition, which increases as the increasing of uniaxial strain along the polarization direction. However, two issues need to be kept in mind. First, the fitted *E-P* curve serves as a guidance only, as in principle the fitting parameters $a$, $b$, and $c$ are temperature dependent which is not reflected in the present DFT calculations carried out at 0K. Second, ideal coercive field $\mathcal{E}_c$ refers to the electric field required to reverse the polarization in a coherent transformation throughout the whole crystal, while in reality true coercive field corresponds to critical electric field that destabilizes the domain walls, which is often much smaller than the ideal coercive field $\mathcal{E}_c$.

2D materials can often sustain large elastic strain (*37, 38*), making it distinctly different from their 3D bulk counterpart. Specifically, the dual ferroic (ferroelastic and ferroelectric) properties in MX monolayers may be subject to continuous alternations upon elastic stress, which may offer a facile control of charge polarization and phase transition barrier and thus provide a broader design space for optoelectronic devices. To verify this concept, we apply a uniaxial tensile strain along *x*-axis (*i.e.* ferroelectric direction) and fully relax the lattice parameter *b* and the atomic positions. The resulted strain-dependent energy profile as function of total polarization is presented in Fig. 3D for monolayer GeSe, manifesting a large effect of elastic strain on the spontaneous polarization. For example, by increasing uniaxial strain from 0% to 6%, the spontaneous polarization in monolayer GeSe is markedly enhanced from 357 $pC/m$ to 430 $pC/m$. The strain-dependent polarization as shown in Fig. 3E also allows us to estimate the piezoelectric coefficient $e_{11} \equiv \frac{\partial P_x}{\partial \epsilon_{xx}}$ of 1.31 nC/m for monolayer GeSe (see Table S7 for other materials), which is in good agreement with the recent work by Fei *et al*. (*13*). Furthermore, as shown in Fig. 3F, the strain-dependent coherent ferroelectric transition energy barrier enlarges from 0.08 to 0.30 eV/unit cell as strain increases from 0% to 6% (fig. S2). The above results unequivocally demonstrate that elastic strain can have significant impact on both spontaneous polarization and coherent transition barriers, highlighting strain modulation as a potential avenue for fine-tuning ferroelectric properties of materials.

The potential energy surfaces shown above illustrate only the thermodynamic properties of perfect crystalline monolayer MX, whereas in reality ferroelectric transition is governed by domain wall motion, analogous to dislocation motion in solids. It is therefore highly important to explore the energetics and transition pathway of ferroelectric domain wall that are relevant to experimental observation and device characterization. In order to acquire a domain wall configuration, we first construct a supercell of pristine monolayer MX consisting of 24 unit cells repeated along the *b* (*y*) axis with ferroelectric polarization along +*x* direction. We then flip the polarization direction of the first 12 unit cells, forming a supercell that contains two 180° ferroelectric domain walls: one at the center (*y=b*/2) and the other at the boundary (*y*=0). Such configuration obviously satisfies the periodic boundary condition required by first-principles DFT using plane wave basis.

An example of 180° ferroelectric domain wall in monolayer GeS is shown in Fig. 4A with its polarization direction indicated by blue and purple arrows. The domain wall energies were then calculated by the energy difference between the fully relaxed supercell and pristine monolayer MX, which yield 116, 56, 24, and 8 meV/Å for GeS, GeSe, SnS, and SnSe, respectively. They correspond to effective domain wall energies of 186, 90, 38, and 13 mJ/m$^2$ if we assume a vdW thickness of 1nm. The effective domain wall energies in monolayer GeS and GeSe are similar to the 180° ferroelectric domain wall energies in prototypic PbTiO$_3$ of 132 and 169 mJ/m$^2$ for Pb-centered and Ti-centered domain wall, respectively, while SnS and SnSe have the domain wall energies that are similar to BaTiO$_3$ with 7.5 and 16.8 mJ/m$^2$ for Ba-centered and Ti-centered domain wall, respectively (*39*).

We then investigate the energy barrier of 180° ferroelectric domain wall migrating along the +*y* direction using first-principles NEB method. Here we focus on monolayer GeS. Its final configuration is obtained by translating the supercell of the initial configuration by one lattice vector along the +*y* axis. Total nine images including the initial and final configurations as well as the linearly interpolated intermediate ones were used in the subsequent NEB calculation. The corresponding MEP is shown in Fig. 4B with the images labeled by 1 to 9. The MEP plot reveals two similar barriers of ~1.6 meV/Å, owing to the fact that the two Ge-S pairs in a single unit cell are related by translational symmetry along the diagonal direction. The resulted three configurations (#1, #5, and #9) with the same lowest energy are shown in fig. S3, where the corresponding domain wall position is marked by dashed line in each configuration. It is important to notice that the migration barrier of ~1.6 meV/Å is remarkably small, which implies that the ferroelectric domain wall in monolayer GeS is highly mobile. Once nucleated, domain wall assisted switching can proceed very fast. We have also calculated MEPs for the other three MX monolayers which are much smaller than that of GeS, indicating that their domain wall motion can be even much faster. It is worth to mention that we also investigated the 90° domain wall, however, the supercell eventually relaxed back to single ferroelectric state due to the large elastic strain energy residing in the supercell. Hence, it is more likely to observe the 180° domain wall discussed above.

In paraelectric state, Ge and S of each local pair exactly overlap on the *x-y* plane, while in the ferroelectric state they are shifted with respect to each other. Therefore, one can extract the domain wall width by calculating the relative displacement of each Ge-S local pair in the supercell. Such relative displacement essentially serves as an order parameter which has two components: one along the *x* direction, and the other along the y direction. Figure 4C shows the corresponding *x* component of relative displacement as a function of the centers of each Ge-S pair, while the *y* component is presented in fig. S3. Thus, according to Fig. 4C, the domain wall width of monolayer GeS is about 1nm which remains almost unchanged for all nine images in the MEP. By comparing the relative M-X displacement for all four materials (fig. S4), we find that GeS has the smallest domain wall width and the largest relative M-X displacement, followed by GeSe, SnS, and SnSe with gradually increased width and reduced displacement, which is consistent with their ground state structure where GeS has the highest spontaneous strain.

The 180° ferroelectric domain wall in monolayer MX is essentially a one-dimensional interface between two ferroelectric states with antiparallel electric polarizations. It is inevitably accompanied by distinct electronic structure localized in the vicinity of the domain wall, which is worth of a detailed study. Figure 4D presents the DFT-PBE band structure of monolayer GeS supercell with 180° ferroelectric domain wall, where different colors indicate the relative localization of electronic states around the domain wall: purple for the states near the domain wall, and cyan for the states away from the domain wall. It clearly shows that the four lowest conduction bands (from the conduction band minimum (CBM) to CBM+3) reside near the domain wall, forming two sets of degenerate bands, *i.e.* degenerate CBM and CBM+1, and degenerate CBM+2 and CBM+3. Figure 4E shows the corresponding real-space wavefunctions for CBM at Γ point, while the wavefunction at CBM+2 can be found in fig. S5. The localized electronic wavefunctions positioned at the low energy conduction bands suggest that they may be easily detected via electronic and optical measurement, for examples, by scanning tunneling microscopy and photoluminescence measurement.

The ferroelectricity and ferroelasticity of monolayer MX discussed above will have more profound impact if they are also coupled to their electronic and optical properties. Figure 5A and fig. S6 show the calculated quasiparticle band structure of MX monolayers where both spin-orbit coupling and quasiparticle effect are included. The quasiparticle effect was taken care by many-body perturbation theory within the $G_0W_0$ approximation (*40, 41*). The results demonstrate that MX monolayers are intrinsically 2D semiconducting materials with indirect band gaps ranging from 2.6 eV to 1.1 eV. We also calculated their photoabsorption spectra by solving two-particle Bethe-Salpeter equation (*42, 43*) based on the quasiparticle energies and the screened Coulomb interactions obtained from the GW calculations. The calculated lowest exciton energies are 2.3, 1.2, 1.8, 1.0 eV for GeS, GeSe, SnS, and SnSe, respectively, whereas their corresponding direct quasiparticle transition energies are 2.9, 1.6, 2.3, and 1.3 eV, resulting in large exciton binding energies of 0.6, 0.4, 0.5, and 0.3 eV that are similar to monolayer $MoS_2$ *etc.* (*38*). In general, the dimensionality reduction leads to a reduced screening in 2D semiconductors, and consequently the effective Coulomb interaction of electron-hole pair becomes much stronger, hence large exciton binding energies in 2D semiconductors are expected upon photoexcitation. Figure 5B presents the calculated imaginary dielectric functions with both excitonic and spin-orbit coupling effect included. It illustrates a highly anisotropic photoabsorption in monolayer GeSe with two giant low-energy excitonic peaks: one at 1.2 eV for the *xx* component and the other at 1.9 eV for the *yy* component of the dielectric tensor. Our results on excitonic behaviors and strong optical absorption agree well with two recent theoretical works (*44, 45*). Similar behaviors are found in the other three materials (fig. S7). Such anisotropic electronic and optical properties are the direct consequences of broken crystalline symmetry due to the ferroelastic spontaneous strain and ferroelectric spontaneous polarization in MX monolayers.

**Discussion**

The coupled ferroelastic and ferroelectric orders and the polarization-dependent optical properties within the visible range imply their great potentials for 2D mechano-opto-electronic device applications by manipulating electrical, mechanical, and optical external fields to control the internal responses. For example, as shown in Fig. 6A, by applying large external electric field or bias voltage one can switch the polarization direction along the same axis, and by measuring the current-voltage curve under small bias one can detect the state without any detrimental effect, thus implementing the "write" and "read" functions of *2D ferroelectric memory (46)*. The corresponding state can also be detected or "read" by measuring the photocurrent upon photo-illumination since the internal electric field will determine the drifting directions of photoexcited charge carriers. Furthermore, owing to their ferroelastic nature, applied mechanical stress can also control the polarization state along different in-plane axes (Fig. 6B), which can be "read" by electrical measurement or by polarization-dependent photoluminescence measurement, rendering a *2D ferroelastic memory*. Moving one step further, one may apply intense coherent light to switch ("write") between the two ferroelastic states through optically-controlled domain wall motion (*47*), and "read" out the states through the polarization-dependent photoluminescence measurement, enabling a *2D ferroelastoelectric non-* *volatile photonic memory* (Fig. 6C). Finally, as shown in Fig. 6D, MX monolayers may serve as an ideal materials platform for realizing *2D ferroelectric photovoltaics*, because (*i*) their optical bandgap falls within the visible range from 1.0 eV to 2.3 eV; (*ii*) their optical absorption is strong due to the pronounced excitonic effect commonly shared by two-dimensional semiconducting materials (Fig. 5B); and (*iii*) the internal in-plane electric field in ferroelectric MX can greatly enhance the separation of the photo-excited charge carriers to form photocurrent.

It is worth to emphasize that the ultrathin two-dimensional nature has another significant implication, that is, the required energy consumption for the "write" and "read" operations in the above 2D ferroelectric/ferroelastic/photonic memory will be much smaller than that in thick films. Since the kinetic barrier of the domain wall motion which controls the switching process is small, the operational speed should be very fast. In addition, the non-volatile photonic memory proposed here is based on the ferroelastic transition between two symmetrically-equivalent crystalline states, which is distinct from the recently-demonstrated first photonic switch based on photo-induced crystalline-to-amorphous phase transition (*48*).

To summarize, in this work we discussed the coupled ferroelastic and ferroelectric orders as well as the electronic, optical, and domain wall properties of monolayer Group IV monochalcogenides. Monolayer MX represents a class of 2D multiferroic semiconductor with large in-plane spontaneous polarization, spontaneous lattice strain, and small domain wall energy, an important addition to the existing realm of multiferroic bulk materials, interface structures, and thin films. The calculated energy barriers of coherent ferroelectric-to-paraelectric and ferroelastic-to-paraelastic transitions show that their dual ferroic orders are thermodynamically stable at room temperature. In addition, the ferroelectric polarization and coherent transition barrier can be easily tuned via elastic strain engineering. The in-plane ferroelectric polarization in 2D MX monolayers makes them distinctly different from ferroelectric thin-films such as perovskite compounds with out-of-plane polarization whose thickness is intrinsically limited by the amplified out-of-plane depolarization effect (*28*). Conversely, a minimal in-plane feature size is expected in MX monolayers due to in-plane depolarization. The strongly-coupled ferroelastic and ferroelectric orders and the polarization-dependent excitonic absorption and photoluminescence within the visible spectrum may enable potential applications for 2D mechano-opto-electronic applications such as 2D ferroelectric excitonic photovoltaics and 2D memory devices including ferroelectric, ferroelastic, and ferroelastoelectric non-volatile photonic memory. The present findings will open up new avenues for miniaturized low-power optoelectronic and photonic applications of 2D multiferroic materials with coupled electronic, optical, mechanical, and even magnetic properties if the symmetry-breaking induced valley-dependent polarization is also considered.

**Materials and Methods**

Atomistic and electronic structures of monolayer Group IV monochalcogenides were calculated using first-principles density-functional theory (DFT) (*29, 30*) as implemented in the Vienna Ab initio Simulation Package (VASP) (*49, 50*). We used the Perdew-Burke-Ernzerhof (PBE) (*51*) form of exchange-correlation functional within the generalized gradient approximation (*52, 53*), plane wave basis with a cutoff energy of 400 eV, and a Monkhorst-Pack (*54*) *k*-point sampling grid of 10×10×1. Ground state structures of monolayer Group IV monochalcogenides were obtained by fully relaxing both atomic positions and in-plane lattice parameters while keeping a large vacuum region along the plane normal to avoid the periodic image interactions. The maximal residual forces for structural relaxation were less than 0.005eV/Å, and the convergence criteria for electronic relaxation is $10^{-6}$ eV. Ferroelectric spontaneous polarization was calculated using the Berry phase approach (*33, 34*), in which both electronic and ionic contributions were taken into account.

To achieve the minimum energy pathways (MEP) of the coherent ferroelastic and ferroelectric transitions as well as the domain wall migration, we applied the climbing image nudged elastic band method (CI-NEB) (*31*) based on the interatomic forces and total energy acquired from the DFT calculations.

Quasiparticle band structures were calculated by many-body perturbation theory calculations within the $G_0W_0$ approximation (*40, 41*). As excitons are expected in 2D semiconductors due to the reduced Coulomb screening, we calculated their dielectric functions and optical absorption spectra by solving two-particle Bethe-Salpeter equation (BSE) (*42, 43*) using the quasiparticle energies, wavefunctions, and screened Coulomb interactions from the $G_0W_0$ calculations. Spin-orbit coupling was included in both $G_0W_0$ and BSE calculations.

**Note added:**

After we completed this work, we learned of a very recent independent work by Wu *et al*. (*55*).

**Supplementary Materials**

fig. S1. Total polarizations along an adiabatic path through the paraelectric state for monolayer GeS, GeSe, SnS, and SnSe.

fig. S2. Coherent ferroelectric transition energy barriers of monolayer GeS, GeSe, SnS, and SnSe under different uniaxial strain.

fig. S3. Ferroelectric domain wall (180°) in monolayer GeS.

fig. S4. Relative displacement of local M-X pairs in monolayer GeS, GeSe, SnS, and SnSe.

fig. S5. Electronic structures of monolayer GeS with 180$^o$ domain wall.

fig. S6. Electronic band structures of monolayer GeS, GeSe, SnS, and SnSe.

fig. S7. Imaginary part of dielectric function in monolayer GeS, GeSe, SnS, and SnSe with excitonic effect and spin-orbit coupling included.

Table S1. Optimized lattice parameters of MX monolayers.

Table S2. Cohesive energy of MX monolayers calculated by DFT-PBE with optB88-vdW.

Table S3. Transformation strain matrices with respect to their paraelastic structure.

Table S4. Coherent ferroelastic transition barrier of MX monolayers at stress-free state.

Table S5. Coherent ferroelectric transition barrier of MX monolayers at strain-free state.

Table S6. Spontaneous polarization of MX monolayers.

Table S7. Piezoelectric coefficients in the large elastic strain limit with up to 6% tensile strain along *x* direction.

**Acknowledgments**

**Funding:** H.W. and X.Q. acknowledge the start-up funds from Texas A&M University and Texas A&M Supercomputing Facility for providing supercomputing resources.

**Author contributions:** X.Q. conceived and supervised the project. H.W. and X.Q. carried out the calculations and analyzed the data. X.Q. wrote the initial draft of manuscript and incorporated the revisions from H.W.

**Competing financial interests:** The authors declare no competing financial interests.

**Data and materials availability:** All data needed to evaluate the conclusions in the paper are present in the paper and/or the Supplementary Materials. Additional data related to this paper may be requested from the authors.

**Figures**

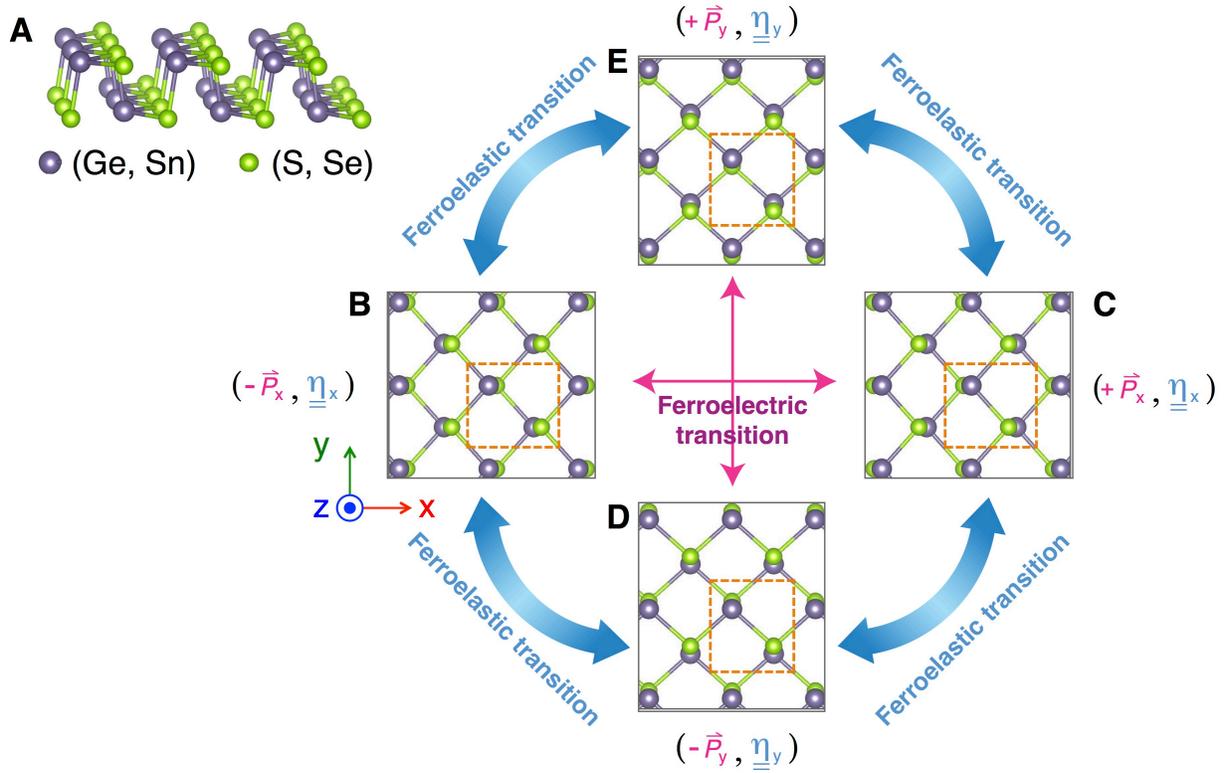

**Fig. 1. Structure of monolayer Group IV monochalcogenides (MX) and their ferroelastic and ferroelectric orders.** (**A**) Perspective view of free-standing MX monolayer using GeSe as an example. (**B**) and (**C**): two ferroelectric states with opposite spontaneous polarization along the $x$ axis: $-P_x$ and $+P_x$. (**D**) and (**E**): two ferroelectric states with opposite spontaneous polarization along the $y$ axis: $-P_y$ and $+P_y$. (**B,C**) and (**D,E**): two ferroelastic states with spontaneous tensile strain along $x$ and $y$ relative to the centrosymmetric paraelastic structure. The corresponding transformation strain matrices are denoted by $\boldsymbol{\eta}_x$ and $\boldsymbol{\eta}_y$.

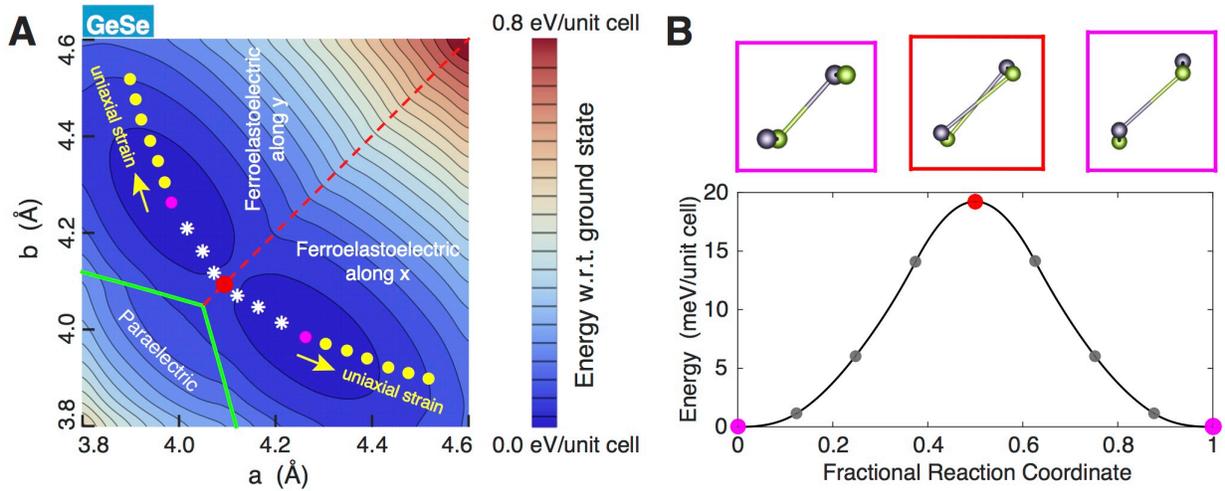

**Fig. 2. Ferroelastic order and spontaneous strain in monolayer GeSe.** (**A**) Total energy of monolayer GeSe with different lattice parameters *a* and *b* with respect to their ferroelectric ground state. Solid green line: phase boundary between paraelectric and ferroelectric phases. Red dashed line: phase boundary between two ferroelastoelectric phases with spontaneous strain along *x* and *y*, respectively, where ferroelectric polarization will be along +*x*/-*x* and +*y*/-*y*, respectively. Two purple dots stand for the two ferroelastoelectric ground states at stress-free condition, while the yellow dots stand for the corresponding lattice parameters upon uniaxial tensile strain along *x* and *y* axis. The contour colors illustrate the total energy of unit cell relative to the energy of ferroelectric ground states (purple dots). (**B**) Minimum energy pathway of coherent ferroelastic transition. The gray dots in (**B**) correspond to the white stars in (**A**).

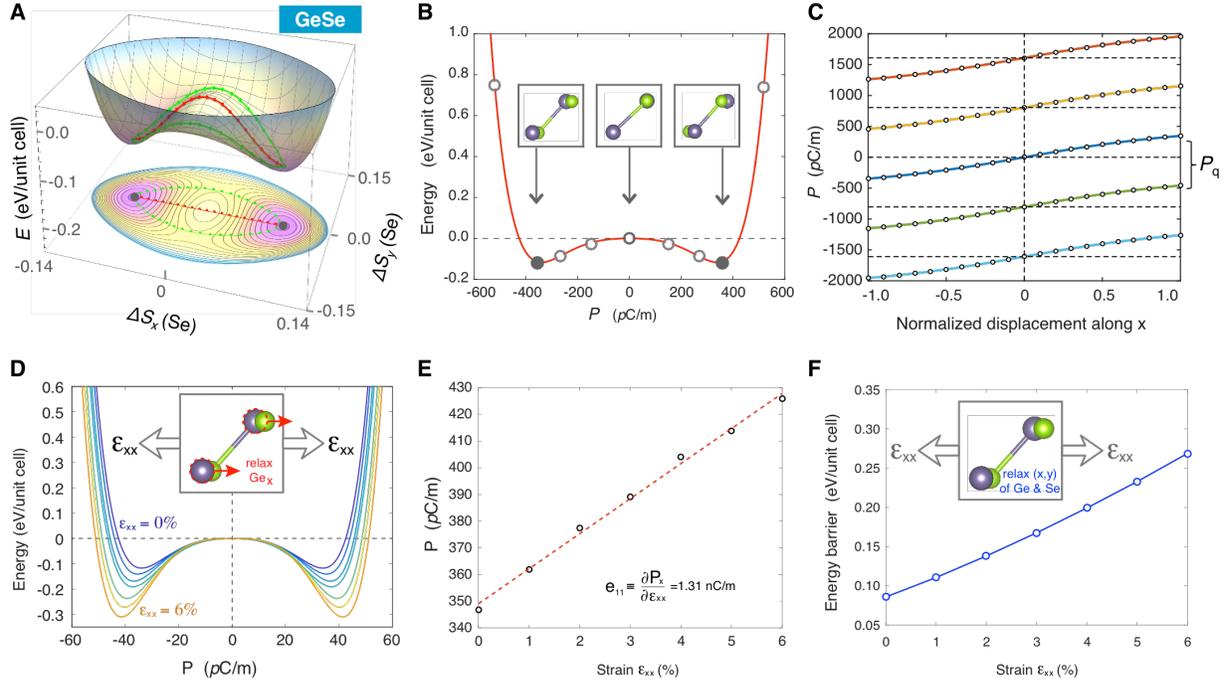

**Fig. 3. Ferroelectric order, spontaneous polarization, and elastic strain effect in monolayer GeSe**. (**A**) Potential energy surface with fractional shift of Se atoms along *x* and *y* with the fixed ground-state lattice parameters. Green lines: MEP for ferroelectric transition. Red line: adiabatic path for polarization calculation. Two black dots: degenerate ferroelectric ground states. (**B**) Double-well potential of monolayer GeSe along the adiabatic path. (**C**) Calculated total polarization as a function of normalized displacement where the centrosymmetric paraelectric phase (0% displacement) is at the center, and two ferroelectric ground states are at two ends. $P_q$ represents the polarization quanta. (**D**) Effect of uniaxial strain on the total energy and total spontaneous polarization. (**E**) Spontaneous polarization of ferroelectric GeSe monolayer as function of uniaxial strain. (**F**) Minimum energy barrier of coherent ferroelectric transition as function of uniaxial strain without the constraint through the paraelectric phase.

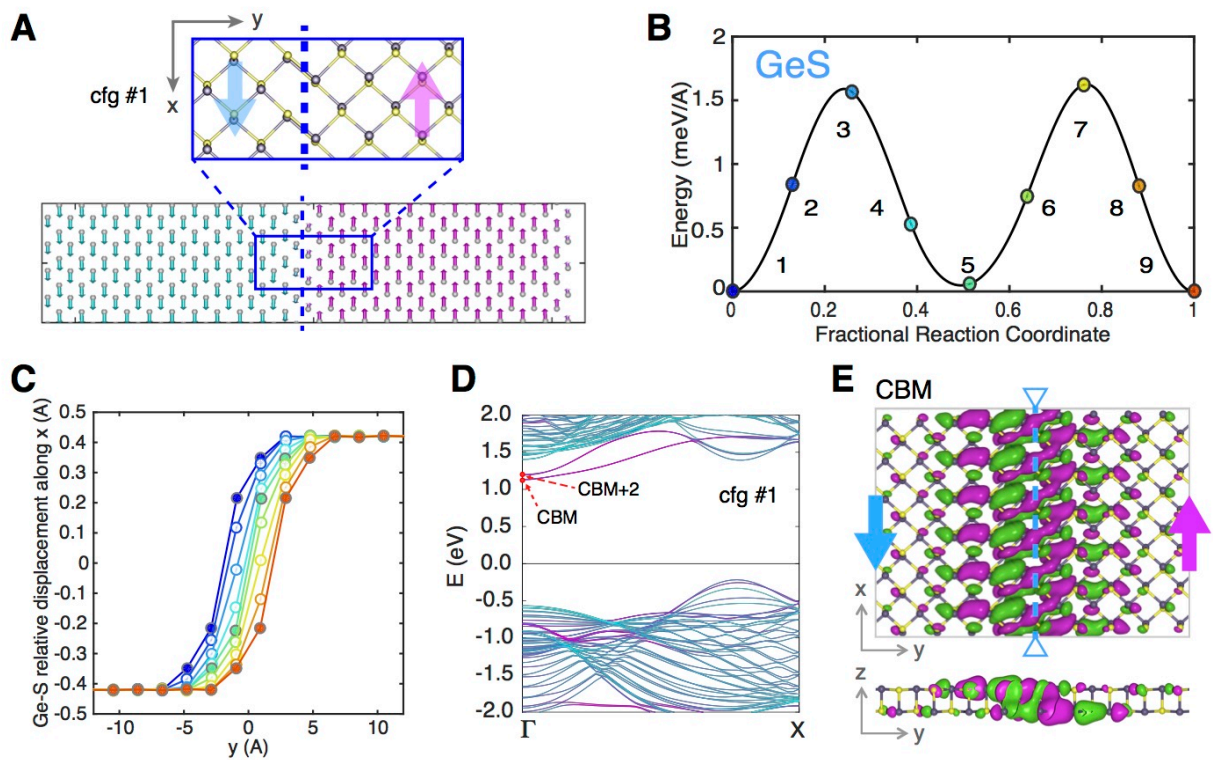

**Fig. 4. Ferroelectric domain wall in monolayer GeS**. (**A**) Ground-state atomistic structure of 180° domain wall. Blue and purple arrows illustrate the polarization direction on the left-hand and right-hand side of the domain wall. (**B**) MEP of domain wall motion with initial, intermediate, and final configurations labeled from 1 to 9. (**C**) Relative displacement of each local Ge-S pair along the x direction. Colors from blue to red correspond to the configurations #1-#9 marked in (**B**). (**D**) Band structure of monolayer GeS supercell with 180° ferroelectric domain wall. Color indicates the relative localization of electronic states around the domain wall. (**E**) Top and side view of Kohn-Sham wavefunction at the conduction band minimum (CBM). Purple and green isosurfaces represent the positive and negative values of wavefunctions, respectively.

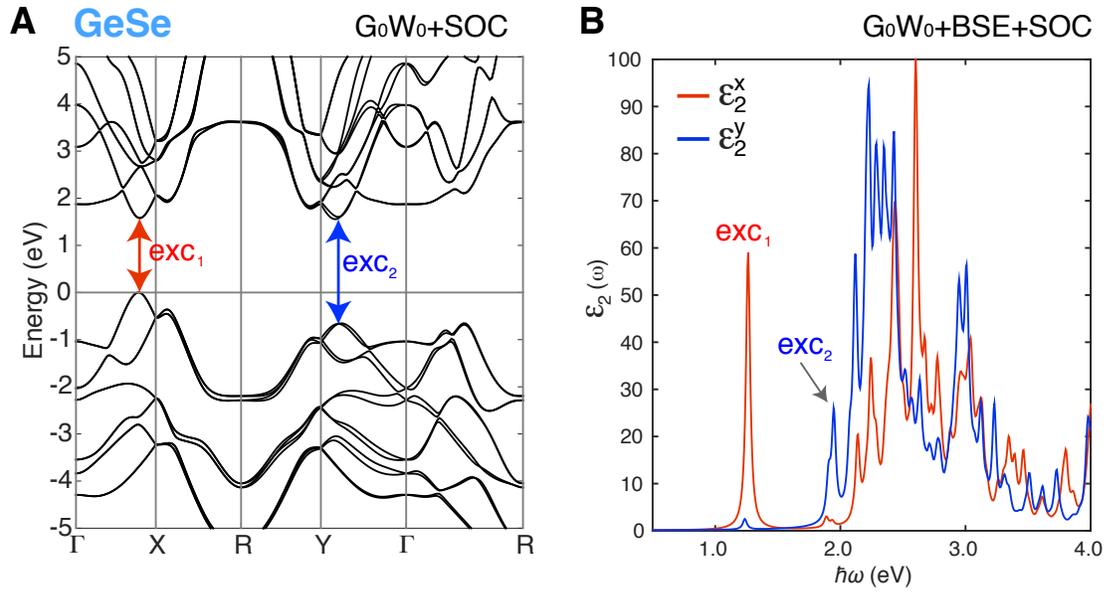

**Fig. 5. Electronic structure of monolayer GeSe**. (**A**) Band structure of pristine monolayer GeSe. (**B**) Imaginary dielectric function along the *x* (red) and *y* (blue) direction. Two arrows in the band structure (**A**) indicate the quasiparticle gap that corresponds to two excitonic peaks in the imaginary dielectric function (**B**).

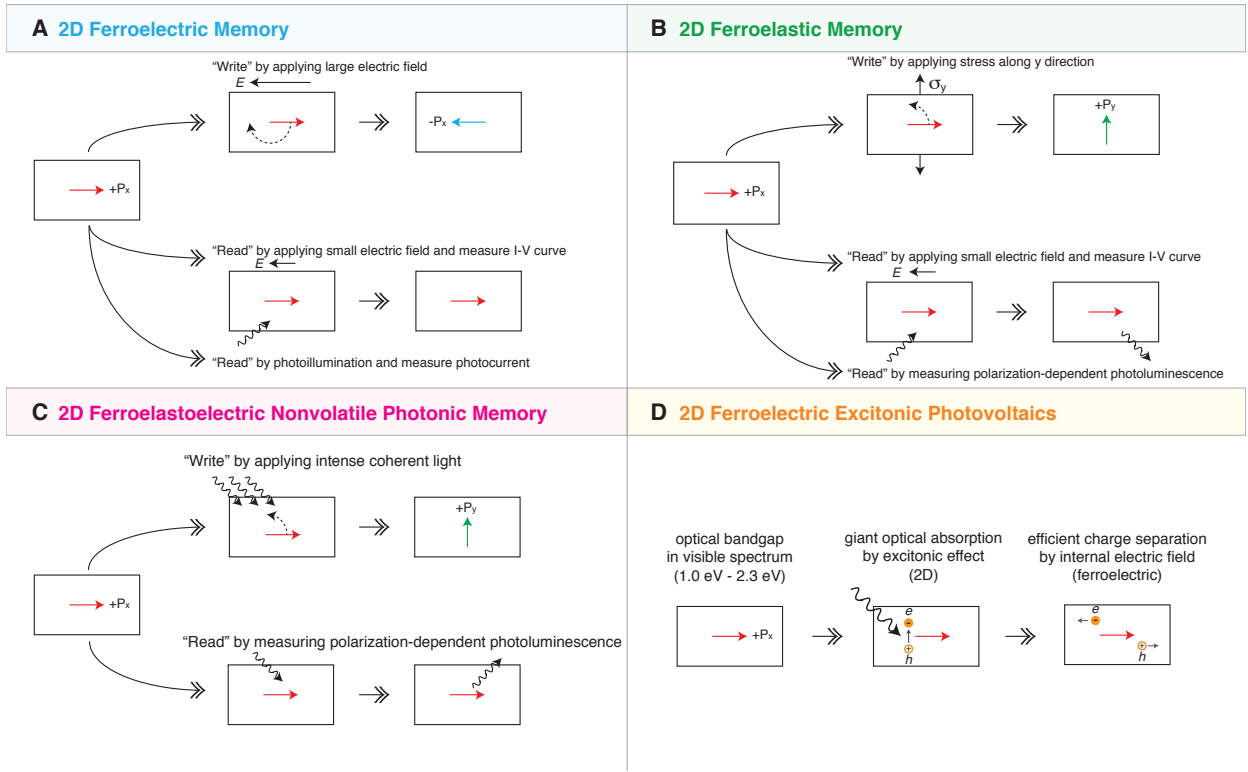

**Fig. 6. Proposed mechano-opto-electronic device concepts and their mechanisms**. (**A**) 2D ferroelectric memory. (**B**) 2D ferroelastic memory. (**C**) 2D ferroelastoelectric nonvolatile photonic memory. (**D**) 2D ferroelectric excitonic photovoltaics.

# Supplementary Materials

# Two-Dimensional Multiferroics: Ferroelasticity, Ferroelectricity, Domain Wall, and Potential Mechano-opto-electronic Applications

Hua Wang[1], Xiaofeng Qian[1]*

[1] *Department of Materials Science and Engineering, Dwight Look College of Engineering and College of Science, Texas A&M University, College Station, Texas 77843, USA*

*Correspondence and requests for materials should be addressed to X.Q.
(email: feng@tamu.edu).

**This file includes:**

fig. S1. Total polarizations along an adiabatic path through the paraelectric state for monolayer GeS, GeSe, SnS, and SnSe.

fig. S2. Coherent ferroelectric transition energy barriers of monolayer GeS, GeSe, SnS, and SnSe under different uniaxial strain.

fig. S3. Ferroelectric domain wall (180°) in monolayer GeS.

fig. S4. Relative displacement of local M-X pairs in monolayer GeS, GeSe, SnS, and SnSe.

fig. S5. Electronic structures of monolayer GeS with 180° domain wall.

fig. S6. Electronic band structures of monolayer GeS, GeSe, SnS, and SnSe.

fig. S7. Imaginary part of dielectric function in monolayer GeS, GeSe, SnS, and SnSe with excitonic effect and spin-orbit coupling included.

Table S1. Optimized lattice parameters of MX monolayers.

Table S2. Cohesive energy of MX monolayers calculated by DFT-PBE with optB88-vdW.

## A. Supplementary Figures

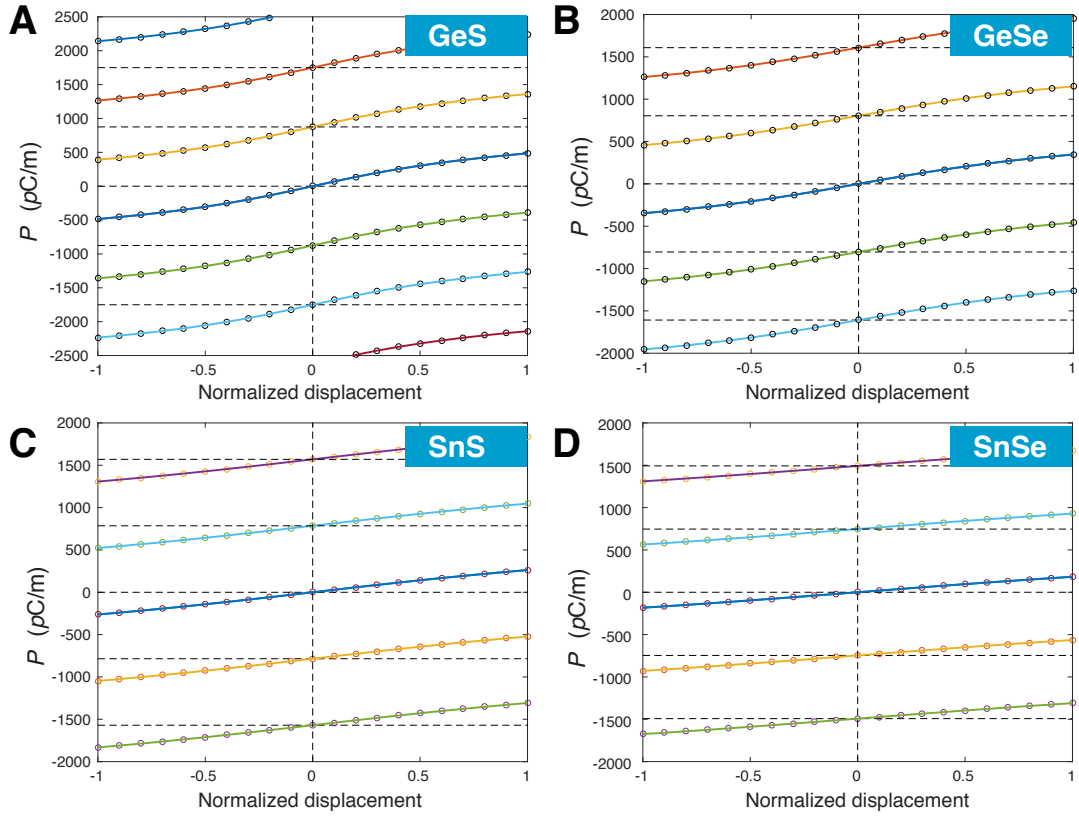

**Fig. S1. Total polarizations along an adiabatic path through the paraelectric state for monolayer GeS, GeSe, SnS, and SnSe.**

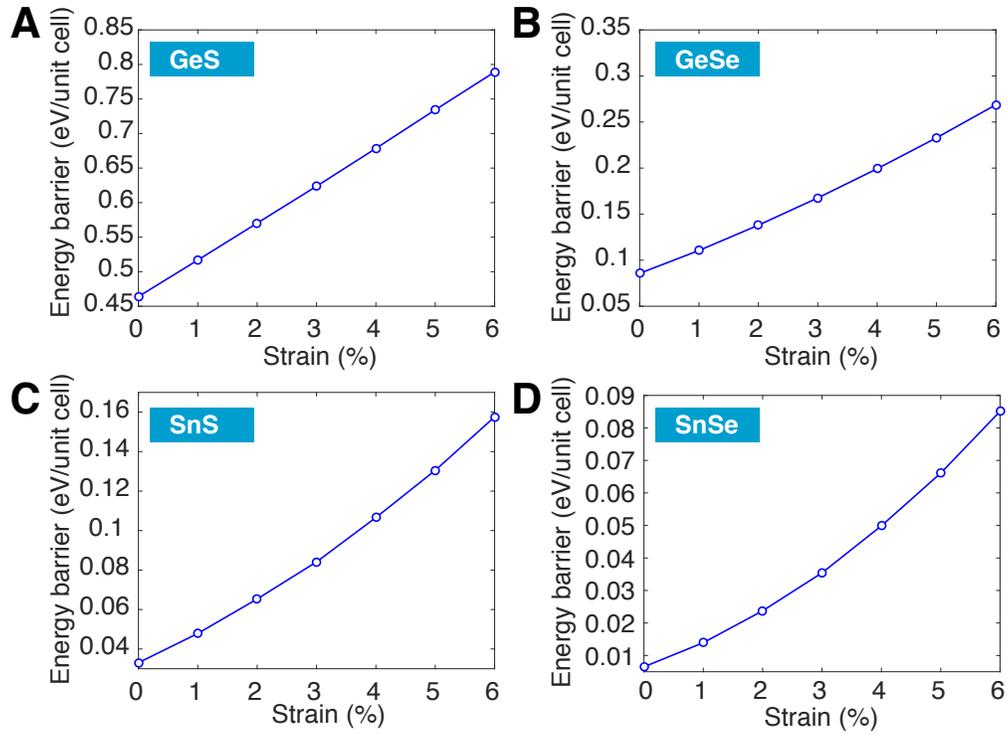

**Fig. S2. Coherent ferroelectric transition energy barriers of monolayer GeS, GeSe, SnS, and SnSe under different uniaxial strain.** The strain is applied along the *x* direction (ferroelectric spontaneous polarization direction) with lattice parameter *b* (along the *y* direction) fully relaxed.

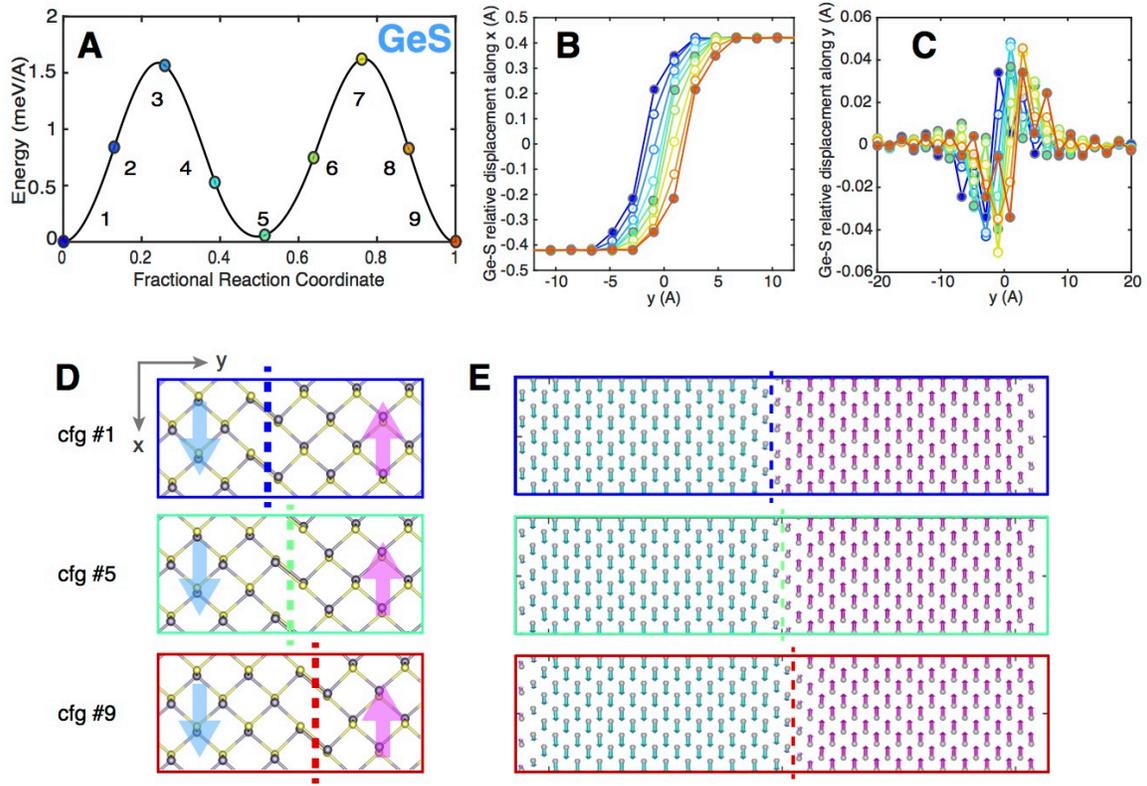

**Fig. S3. Ferroelectric domain wall (180°) in monolayer GeS.** (**A**) Minimum energy pathway of domain wall motion with initial, intermediate, and final configurations labeled from 1 to 9. (**B**) Atomic structures of the initial (#1), middle (#5), and final (#9) configuration. Blue and purple arrows illustrate the polarization direction of the left-hand and right-hand side of domain wall in monolayer GeS, respectively. (**C,D**) Relative displacement of each Ge-S pair along +$x$ and +$y$ direction, respectively. Colors from blue to red refer to the configurations #1-#9, corresponding to the colors of the dots marked in (**A**). (**E**) 2D plots of relative displacement of each Ge-S pair for the configurations #1, #5, and #9 in the supercell. Small blue and purple arrows illustrate the relative displacement direction of each Ge-S pair. Gray spheres stand for the centers of each Ge-S pair.

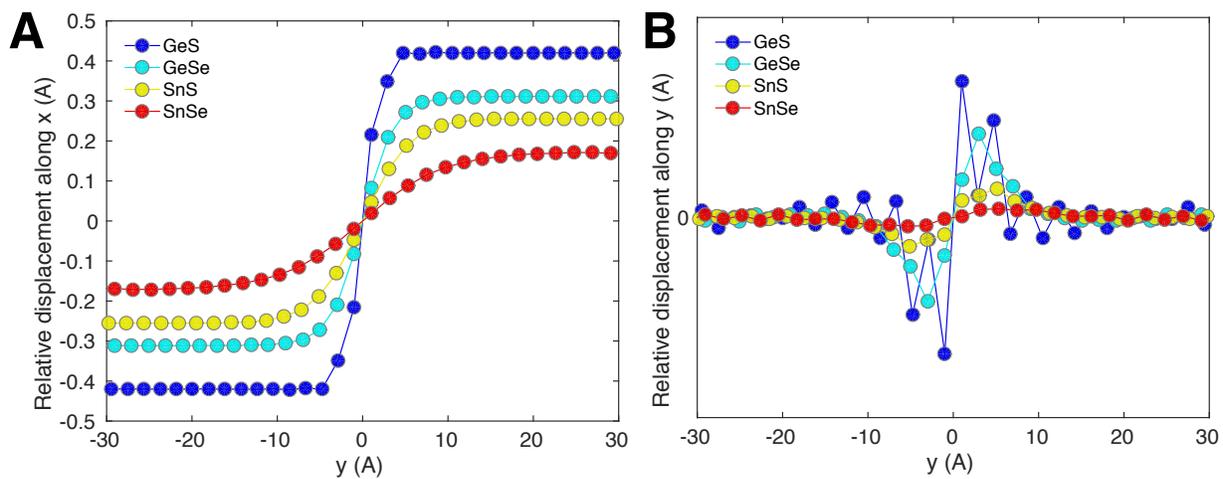

**Fig. S4. Relative displacement of local M-X pairs in monolayer GeS, GeSe, SnS, and SnSe.**
**(A)** Relative displacement along the x direction. **(B)** Relative displacement along the y direction.

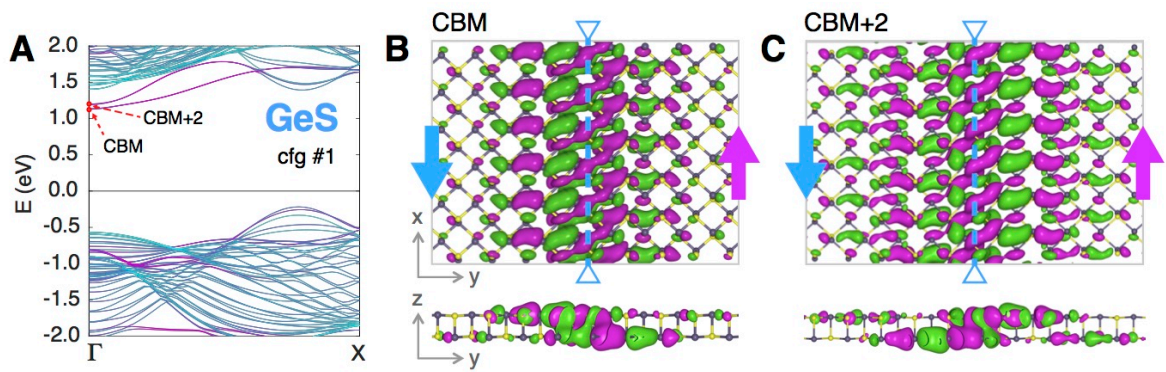

**Fig. S5. Electronic structures of monolayer GeS with 180° domain wall.** (**A**) Electronic band structure. (**B**) Top and front view of the Kohn-Sham wavefunction at the conduction band minimum (CBM). (**C**) Top and front view of the Kohn-Sham wavefunction at the CBM+2. Domain wall is illustrated by dashed blue line. Blue and purple arrows indicate the polarization direction on the left and right hand side of the domain wall.

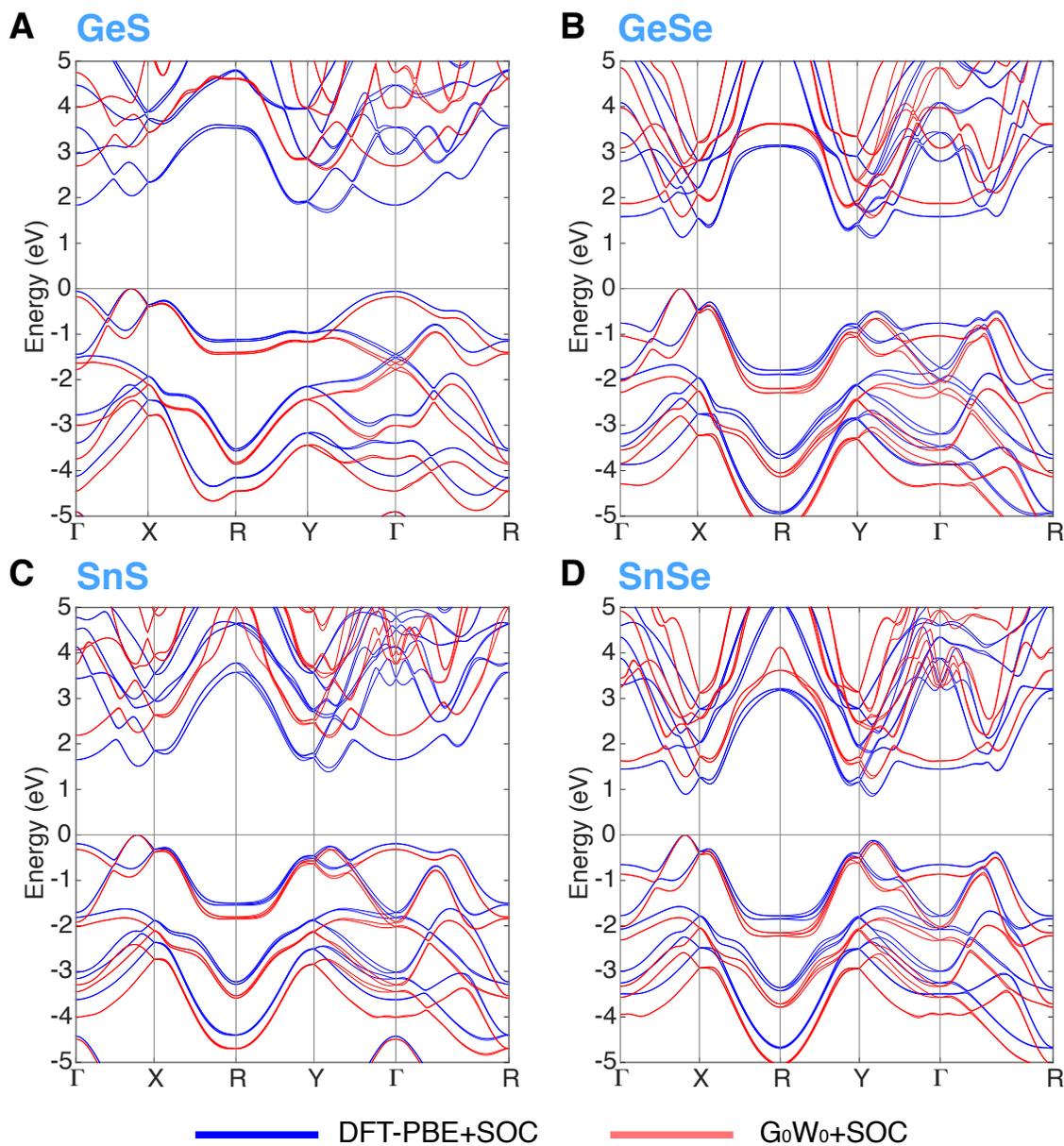

**Fig. S6. Electronic band structures of monolayer GeS, GeSe, SnS, and SnSe.** Blue lines: calculations using DFT-PBE with spin-orbit coupling included. Red lines: calculations using many-body perturbation theory within the $G_0W_0$ approximation with the spin-orbit coupling included as well.

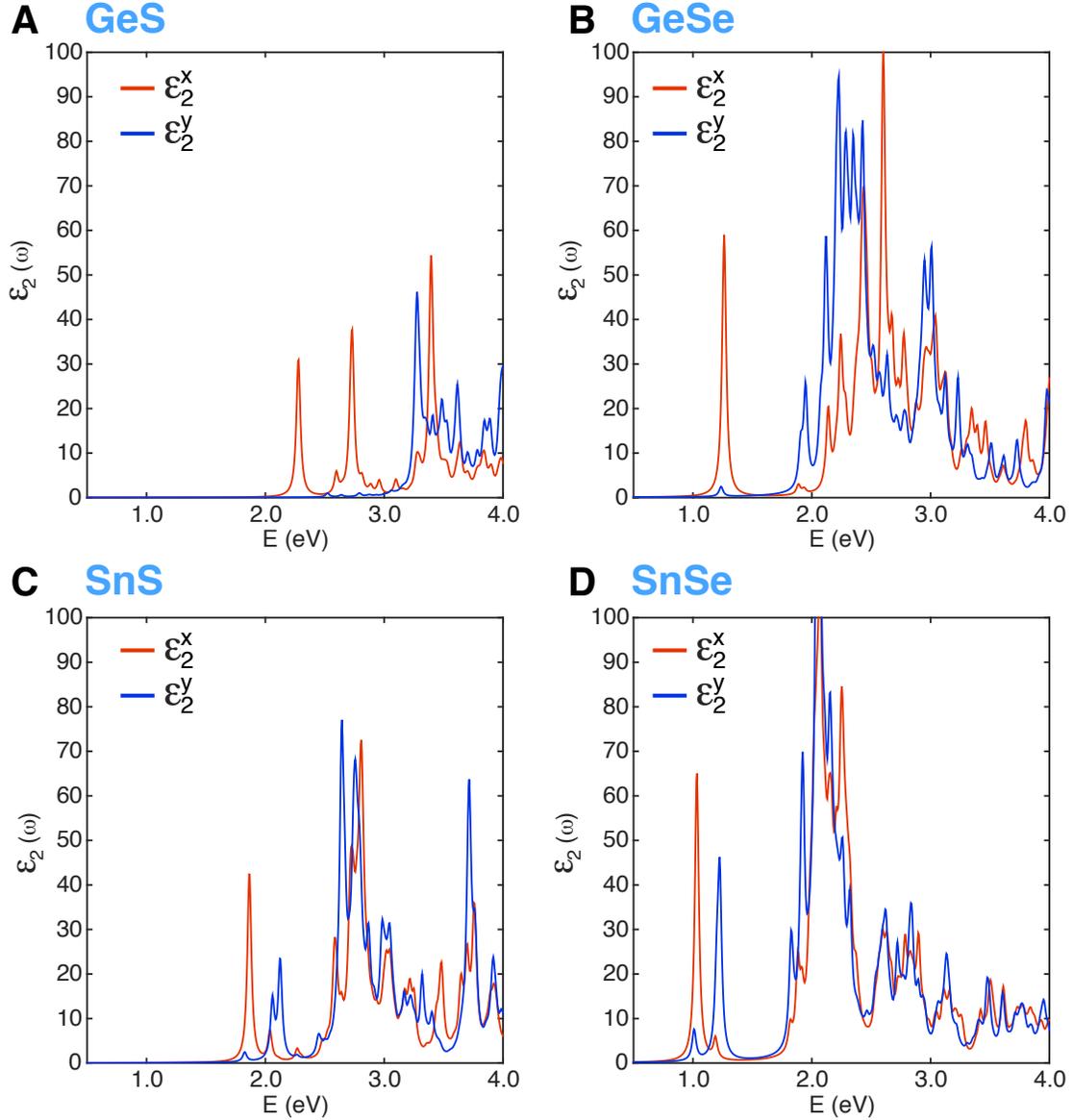

**Fig. S7. Imaginary part of dielectric function in monolayer GeS, GeSe, SnS, and SnSe with excitonic effect and spin-orbit coupling included.** Red lines: the *xx* component of imaginary dielectric tensor. Blue lines: the *yy* component of imaginary dielectric tensor. Calculations are done by solving the two-particle Bethe-Salpeter equation using quasiparticle energies and screened Coulomb interaction obtained from many-body perturbation theory within the $G_0W_0$ approximation.

## B. Supplementary Tables

| Materials | Monolayer (DFT calculations) | | Bulk (experiments)[1-4] | |
|---|---|---|---|---|
| | $a$ (Å) | $b$ (Å) | $a$ (Å) | $b$ (Å) |
| GeS | 4.47 | 3.66 | 4.30 | 3.64 |
| GeSe | 4.26 | 3.98 | 4.38 | 3.82 |
| SnS | 4.28 | 4.08 | 4.33 | 3.99 |
| SnSe | 4.38 | 4.30 | 4.44 | 4.14 |

**Table S1. Optimized lattice parameters of MX monolayers**. Vacuum regions of 15 Å were used to avoid self-interaction between the neighboring images. Length is in the unit of Å.

| Materials | GeS | GeSe | SnS | SnSe |
|---|---|---|---|---|
| Cohesive Energy (meV/Å$^2$) | 34.76 | 34.29 | 33.40 | 32.72 |

**Table S2. Cohesive energy of MX monolayers calculated by DFT-PBE with optB88-vdW**.

| Materials | $\eta_x$ | $\eta_y$ |
|---|---|---|
| GeS | $\begin{bmatrix} 0.141 & 0 \\ 0 & -0.070 \end{bmatrix}$ | $\begin{bmatrix} -0.070 & 0 \\ 0 & 0.141 \end{bmatrix}$ |
| GeSe | $\begin{bmatrix} 0.041 & 0 \\ 0 & -0.027 \end{bmatrix}$ | $\begin{bmatrix} -0.027 & 0 \\ 0 & 0.041 \end{bmatrix}$ |
| SnS | $\begin{bmatrix} 0.028 & 0 \\ 0 & -0.019 \end{bmatrix}$ | $\begin{bmatrix} -0.019 & 0 \\ 0 & 0.028 \end{bmatrix}$ |
| SnSe | $\begin{bmatrix} 0.010 & 0 \\ 0 & -0.009 \end{bmatrix}$ | $\begin{bmatrix} -0.009 & 0 \\ 0 & 0.010 \end{bmatrix}$ |

**Table S3. Transformation strain matrices with respect to their paraelastic state**.

| Materials | GeS | GeSe | SnS | SnSe |
|---|---|---|---|---|
| Transition barriers (meV/unit cell) | 56.44 | 19.23 | 8.56 | 2.76 |

**Table S4. Coherent ferroelastic transition barrier of MX monolayers at stress-free state using generalized solid-state nudged elastic band method**.

| Materials | GeS | GeSe | SnS | SnSe |
|---|---|---|---|---|
| Transition barriers (meV/unit cell) | 463.99 | 95.34 | 33.05 | 6.51 |

**Table S5. Coherent ferroelectric transition barrier of MX monolayers calculated using the nudged elastic band method with the ground-state unit cell (*i.e.* at strain-free state).**

| Materials | GeS | GeSe | SnS | SnSe |
|---|---|---|---|---|
| Spontaneous Polarization ($p$C/m) | 484 | 357 | 260 | 181 |
| Effective Spontaneous Polarization ($\mu$C/cm$^2$) | 48.4 | 35.7 | 26.0 | 18.1 |

**Table S6. Spontaneous polarization of MX monolayers**. A vdW thickness of 1nm is assumed when estimating effective spontaneous polarization.

| Materials | GeS | GeSe | SnS | SnSe |
|---|---|---|---|---|
| $e_{11}$ (nC/m) | 0.84 | 1.31 | 1.62 | 2.07 |

**Table S7. Piezoelectric coefficients in the large elastic strain limit with up to 6% tensile strain along *x* direction**.